# Exact Solution and Correlation Functions of Generalized Double Ising Chains

P. V. Khrapov, S. A. Shchurenkov

*Abstract*—In this paper the exact solution and correlation functions for a double-chain Ising model with multi-spin interactions and symmetric Hamiltonian density are obtained.
   The study employs the transfer matrix method to derive fundamental thermodynamic characteristics of the system. The main results include exact expressions for the partition function, free energy, internal energy, specific heat capacity, magnetization, susceptibility, and entropy in a strip of finite length and in the thermodynamic limit. The work provides explicit formulas for the eigenvalues and shows structure of eigenvectors of the transfer matrix. The expression for magnetization in the thermodynamic limit using components of normalized eigenvector corresponding to the maximum eigenvalue is obtained.
   A detailed analysis is conducted for a special case of interactions involving all kinds of two- and four-spin interactions. This gives the simplified formula for free energy, it is calculated using the root of quadratic equation. Using special relation between parameters the solution for gonihedric model on the plane is obtained. The research reveals properties of the system, including specific features of ground states and phase diagram characteristics. Particular attention is given to the behavior of physical quantities near frustration points and the investigation of spin correlation functions. Plots of physical characteristics, including inverse correlation length, illustrating the obtained results are constructed.

## I. Introduction

For many years, researchers have been studying magnetic systems and their properties. The Ising model, introduced in 1925, was and remains important for understanding both magnetism and other physical phenomena [1]. The two-dimensional model was solved by Onsager [2] for the interaction of the nearest neighbors.

The two-dimensional Ising model with uncrossed second-neighbor interactions was solved exactly [3]. In this situation, it is of interest to study a lower-dimensional model for which accurate results can be obtained.

In the work of L. Kalok and L. C. de Menezes [4] a system with crossed interactions is studied - a double chain of spins with different exchange interaction constants between the nearest neighbors along each chain and between chains, as well as with an additional crossed interaction of the second neighbors. Such a double chain demonstrates strong frustration, leading to almost degenerate ground states with comparable values of competing interaction constants. In particular, the phase diagram of the ground state (in the absence of a field) includes four regions separated by the boundaries of "compensation points" [4] ("frustration points"), at which frustrating behavior is observed on the graphs of specific heat, inverse correlation length, etc.

In the work of T. Yokota [5] an exact solution and correlation functions for the generalized three chain model enclosed in both directions are obtained with the same Hamiltonian density with nearest and next-nearest interactions as used in [4]

In [6] an exact solution and correlation functions are described for generalized three-chain Ising model with arbitrary multi-spin interactions and Hamiltonian density invariant with respect to global spin shifts across all three-spin layers.

To investigate such models, researchers employ both numerical and analytical approaches. For exact analytical solutions, the transfer matrix method, first introduced in [7,8] and the combinatorial method [9], are used for analytical solutions. Among numerical techniques, variants of the Monte Carlo method and the Metropolis algorithm predominate [10].

Paper [11] is focused on the investigation of stochastic operator spectra. Meanwhile, the cluster decomposition technique outlined in [12] has significantly advanced the research of lattice models.

In [13] the cluster decomposition approach is demonstrated and the transfer matrix spectrum for the two-dimensional Ising model under a strong external field is examined.

The article [14] presents an analysis of phase diagrams for a cubic lattice incorporating both nearest-neighbor and next-nearest-neighbor interactions.

In recent years there have been renewed interest in Ising-type spin systems following the examination of the random surface model within the framework of string theory [16].

Beyond conventional approaches, novel method for solving such models continue to emerge. As an illustration, [17] explores a cubic lattice model featuring nearest-neighbor, next-nearest-neighbor, and plaquette interactions through the application of the cluster variation method.

The classical Hopfield network [18] with binary neurons ($\pm 1$) and symmetric weights is mathematically equivalent to the generalized Ising model with pairwise interactions on a fully connected graph (i.e. the spin glass model). In this equivalence, each neuron corresponds to an Ising spin ($\pm 1$), and the symmetric synaptic weights $w_{ij}$ play the role of the

P. V. Khrapov – Bauman Moscow State Technical University (5/1 2-nd Baumanskaya St., Moscow 105005, Russia),
ORCID: https://orcid.org/0000-0002-6269-0727 ,
e-mail: pvkhrapov@gmail.com , khrapov@bmstu.ru.
S. A. Shchurenkov – Bauman Moscow State Technical University (5/1 2-nd Baumanskaya St., Moscow 105005, Russia),
ORCID: https://orcid.org/0009-0002-4769-3855,
e-mail: schurenkovsa@student.bmstu.ru , stepan.shchurenkov201@gmail.com.

interaction coefficients J_ij between pairs of spins. The energy function of the Hopfield network is identical to the Ising Hamiltonian, meaning that the asynchronous dynamics of the network update emulates the relaxation of the Ising system at zero temperature, minimizing its energy—transitioning to stable attractor states corresponding to the learned patterns.

Nezhadhaghighi [19] investigates the critical scaling and conformal invariance of the Baxter–Wu model (a triangular lattice Ising model with three-spin interactions on each face) at its critical point, using finite-size scaling and conformal field theory methods to confirm the model's critical exponents and its conformal properties, thereby firmly establishing that this multi-spin interaction model belongs to a universality class distinct from the standard Ising model.

In [20] derives an exact solution for the generalized two-dimensional Ising model in a field with nearest-nearest-neighbor, next-nearest-neighbor, ternary, and quadruple spin interactions by constructing a transfer matrix with a special eigenvector, such that the largest eigenvalue remains constant on a certain manifold of coupling parameters (the "disorder solution"), thereby obtaining expressions in analytical form for the free energy in the thermodynamic limit.

Osabutei [21] investigates the mean-field Ising model extended by three-spin interaction, revealing a complex phase diagram with two distinct coexistence curves and two second-order phase transition points, and showing that the critical exponents remain consistent with the mean-field universality class.

Suzuki [22] analyzes spin-$S$ Ising models with $p$-spin interactions (including up to $p=5$ spins simultaneously) on one- and two-dimensional lattices using transfer matrix methods and simulations to demonstrate that higher-order interactions significantly enhance spin correlations (especially at low temperatures for $S \geq 1$) and can lead to stronger first-order character in the finite-temperature phase transition.

In this paper, we consider the general symmetric form of the Hamiltonian density for a double chain Ising model, introducing, in addition to Kalok and de Menezes [4], the field action, triple and quadruple interactions. Using the transfer matrix method, we obtain an exact solution to this model, that is, the partition function, free energy, internal energy, specific heat, magnetization, susceptibility, and entropy. The theorems proved in the paper are relevant both for the general (considered in the first part of the paper) case and for more specific cases. Our goal is to find out what properties a model with interactions of only an even number of spins has - the thermodynamic properties and the phase diagram of this system.

Section 2 provides the derivation of exact expressions for the thermodynamic functions of the generalized double chain Ising model for arbitrary values of the constants $H, J_1, J_2, J_3, J_4$ and $J_5$ for the Ising model.

Section 3 examines the free energy, internal energy, heat capacity, and spin-spin correlation function for spins at the same and different levels in a special case of interactions of even number of spins. In addition, a phase diagram of the ground states is presented (Figure 2) showing several regions, separated by lines of "compensation points". Illustration of physical characteristics, including inverse correlation length, are also shown in this section. Regions of ground states, which were found according to the work [23], are introduced graphically in three-dimensional space for the case $J_1>0$.

Section 4 presents the proofs of the theorems and the methods to find the eigenvalues and eigenvectors of the transfer matrix.

An exact solution of the double-chain Ising model with an external field, as well as interactions of two, three, and four spins, is obtained.

## II. MODEL DESCRIPTION AND MAIN RESULTS

### A. Model Description

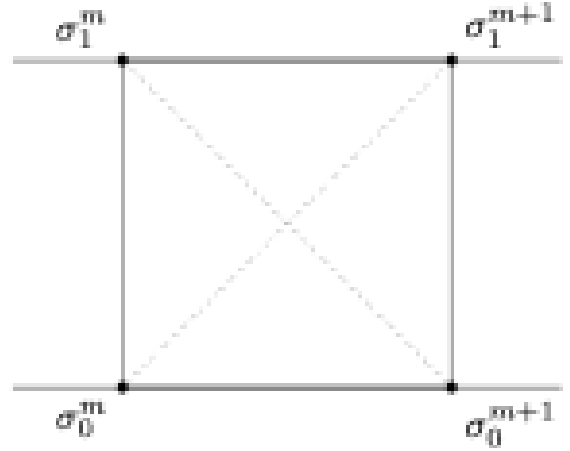

*Figure 1. Configuration of two-chain open model*

For the cyclically closed two-chain Ising model of length $L$ we will consider a symmetric Hamiltonian density invariant under substitutions $\sigma_i^m \leftrightarrow \sigma_i^{m+1}, \sigma_i^m \leftrightarrow \sigma_{i+1}^m$, as well as their composition. Hamiltonian density is

$$\mathcal{H}^m = -J_1\left(\sigma_0^m \sigma_0^{m+1} + \sigma_1^m \sigma_1^{m+1}\right) - \frac{J_2}{2}\left(\sigma_0^m \sigma_1^m + \sigma_0^{m+1} \sigma_1^{m+1}\right) -$$
$$- J_3\left(\sigma_0^m \sigma_1^{m+1} + \sigma_0^{m+1} \sigma_1^m\right) - J_4 \sigma_0^m \sigma_0^{m+1} \sigma_1^m \sigma_1^{m+1} -$$
$$- J_5\left(\sigma_0^m \sigma_0^{m+1} \sigma_1^m + \sigma_0^m \sigma_0^{m+1} \sigma_1^{m+1} + \sigma_0^m \sigma_1^{m+1} \sigma_1^m + \sigma_0^{m+1} \sigma_1^{m+1} \sigma_1^m\right) -$$
$$- \frac{H}{2}\left(\sigma_0^m + \sigma_1^m + \sigma_0^{m+1} + \sigma_1^{m+1}\right), \quad (1)$$

$$\mathcal{H} = \sum_{m=0}^{L-1} \mathcal{H}^m. \quad (2)$$

Since interactions between the elements of carriers $\{\sigma_0^m \sigma_1^m\}, \{\sigma_0^{m+1} \sigma_1^{m+1}\}$ and $\{\sigma_0^m\}, \{\sigma_1^m\}, \{\sigma_0^{m+1}\}, \{\sigma_1^{m+1}\}$ occur twice, we will take into account the factor ½ in the sums associated with these interactions (1).

From now on, such units of measurement thar Boltzmann constant equals one will be used.

### B. Partition function

The most important quantity considered in statistical mechanics is the partition function, knowing which it is possible to obtain the exact solution of the model. Its formula is (where the summation is carried out over all possible sets $\{\sigma\}$):

$$Z_L = \sum_{\{\sigma\}} exp\left\{-\frac{1}{T}\mathcal{H}(\sigma)\right\} \quad (3)$$

To calculate partition function the transfer matrix is introduced:

$$\Theta = \begin{pmatrix} a & b & b & c \\ b & d & f & g \\ b & f & d & g \\ c & g & g & h \end{pmatrix}, \quad (4)$$

$$\theta_{i,j} = e^{-\mathcal{H}(\sigma)}. \quad (5)$$

This can be show schematically as:

*Table 1. The principle of transfer matrix construction*

|  |  | $\sigma_0^{m+1}$ | + | + | − | − |
|---|---|---|---|---|---|---|
|  |  | $\sigma_1^{m+1}$ | + | − | + | − |
| $\sigma_0^m$ | $\sigma_1^m$ |  |  |  |  |  |
| + | + |  | a | b | b | c |
| + | − |  | b | d | f | g |
| − | + |  | b | f | d | g |
| − | − |  | c | g | g | h |

**Theorem 1.** *The partition function for two-chain Ising model of length L can be written as*
$Z_L = \lambda_1^L + \lambda_2^L + \lambda_3^L + \lambda_4^L,$
*where $\lambda_1, \lambda_2, \lambda_3, \lambda_4$ are eigenvalues of transfer matrix (4). One of them can be easily found, we will call it $\lambda_4$, it is equal to d−f. To find three other eigenvalues one needs to solve cubic equation:*
$\lambda^3 + F\lambda^2 + G\lambda + S = 0,$
*where*
$F = -a - d - f - h,$
$G = -2b^2 - c^2 + ad + af - 2g^2 + ah + dh + fh,$
$S = c^2 d + c^2 f - 4bcg + 2ag^2 + 2b^2 h - adh - afh.$
*Then eigenvalues are found using formulas:*

$$\lambda_k = \frac{1}{3}\left(-F + 2\sqrt{F^2 - 3G}\sin\left[\frac{1}{3}\left(\arcsin\left[\frac{2F^3 - 9FG + 27S}{2(F^2 - 3G)^{1.5}}\right] + 2\pi k\right)\right]\right),$$

$k = 1, 2, 3,$

$\lambda_4 = d - f.$

*Where $\lambda_1 = \lambda_{max}$, $\lambda_{max}$ is characterized as positive and has the greatest absolute value among $\lambda_i$.*

By using properties of permutation matrices, we factorize characteristic polynomial of matrix (4) and find its eigenvalues (section IV).

We obtain the following expressions for free energy [24], internal energy, specific heat capacity, magnetization, susceptibility and entropy respectively [6]:

$$f(H,T) = -\frac{T}{2L}\ln(Z_L(H,T)), \quad (6)$$

$$u(H,T) = -T^2\frac{\partial}{\partial T}\left[\frac{f(H,T)}{T}\right], \quad (7)$$

$$c(H,T) = \frac{\partial}{\partial T}u(H,T), \quad (8)$$

$$M(H,T) = -\frac{\partial}{\partial H}f(H,T), \quad (9)$$

$$\chi(H,T) = \frac{\partial}{\partial H}M(H,T), \quad (10)$$

$$S(H,T) = -\frac{\partial}{\partial T}f(H,T). \quad (11)$$

Now we can formulate the main theorem.

**Theorem 2 (The main theorem).** *In the thermodynamic limit for the open two-chain model with Hamiltonian (1) free energy, internal energy, specific heat capacity, magnetization, susceptibility and entropy (6)-(11), respectively, are calculated using the following expressions:*

$$f(H,T) = -\frac{T}{2}\ln(\lambda_{max}), \quad (12)$$

$$u(H,T) = T^2\frac{\partial}{\partial T}\left[\frac{1}{2}\ln(\lambda_{max})\right], \quad (13)$$

$$c(H,T) = 2T\frac{\partial}{\partial T}\frac{1}{2}\ln(\lambda_{max}) + T^2\frac{\partial^2}{\partial T^2}\frac{1}{2}\ln(\lambda_{max}), \quad (14)$$

$$M(H,T) = \frac{T}{2\lambda_{max}}\frac{\partial \lambda_{max}}{\partial H}, \quad (15)$$

$$\chi(H,T) = \frac{T}{2}\left[\frac{1}{\lambda_{max}}\frac{\partial^2 \lambda_{max}}{\partial H^2} - \left(\frac{1}{\lambda_{max}}\frac{\partial \lambda_{max}}{\partial H}\right)^2\right], \quad (16)$$

$$S(H,T) = \frac{1}{2}\left[\ln(\lambda_{max}) + \frac{T}{\lambda_{max}}\frac{\partial \lambda_{max}}{\partial T}\right]. \quad (17)$$

In paper [25] formulas to calculate partial derivatives in formulas (12)-(17) using coefficients of characteristic polynomial of matrix (4) have been formulated.

## C. Correlation functions and correlators

The diagonalizing matrix for matrix (4):

$$Q = \begin{pmatrix} \frac{\alpha_1}{N_1} & \frac{\alpha_2}{N_2} & \frac{\alpha_3}{N_3} & 0 \\ -\frac{\beta_1}{N_1} & -\frac{\beta_2}{N_2} & -\frac{\beta_3}{N_3} & \frac{1}{\sqrt{2}} \\ -\frac{\beta_1}{N_1} & -\frac{\beta_2}{N_2} & -\frac{\beta_3}{N_3} & -\frac{1}{\sqrt{2}} \\ \frac{\gamma_1}{N_1} & \frac{\gamma_2}{N_2} & \frac{\gamma_3}{N_3} & 0 \end{pmatrix}, \quad (18)$$

In case $\lambda_2 \neq \lambda_3$:

$$\alpha_i = \frac{2g^2 - (d+f)h + (d+f+h)\lambda_i - \lambda_i^2}{cd + cf - 2bg - c\lambda_i}, \quad (19)$$

$$\beta_i = \frac{cg - bh + b\lambda_i}{cd + cf - 2bg - c\lambda_i},$$

$\gamma_i = 1,$

$i = 1, 2, 3.$

In case $\lambda_2 = \lambda_3$:

$\alpha_2 = 0, \beta_2 = 1, \gamma_2 = -\frac{2b}{c},$

$\alpha_3 = -\frac{2b}{a - \lambda_3}, \beta_3 = \frac{2b^2}{c^2 + 2b^2}, \gamma_3 = \frac{2bc}{c^2 + 2b^2}.$

Eigenvectors are normalized with coefficients:

$N_i = \sqrt{\alpha_i^2 + 2\beta_i^2 + \gamma_i^2}, \; i = 1, 2, 3.$

As a result, expressions (18) and (19) deliver a detailed exposition of the eigenvector form. Now we can formulate the following theorem.

**Theorem 3.** *For two-chain open model where $L \to \infty$: The average spin value of $\sigma_i$:*

$$\langle \sigma_i \rangle = \frac{1}{N_1^2}(\alpha_1^2 - \gamma_1^2).$$

*If the spins are on the same chain at a distance of $k = |j-i|$:*

$$\langle\sigma_i\sigma_j\rangle = 2\frac{\beta_1^2}{N_1^2}\left(\frac{\lambda_4}{\lambda_1}\right)^k + \frac{(\alpha_1\alpha_2 - \gamma_1\gamma_2)^2}{N_1^2 N_2^2}\left(\frac{\lambda_2}{\lambda_1}\right)^k +$$

$$+ \frac{(\alpha_1\alpha_3 - \gamma_1\gamma_3)^2}{N_1^2 N_3^2}\left(\frac{\lambda_3}{\lambda_1}\right)^k + \frac{(\alpha_1^2 - \gamma_1^2)^2}{N_1^4}.$$

*If the spins are on different chains at a distance of $k=|j-i|$:*

$$\langle\sigma_i\sigma_j\rangle = -2\frac{\beta_1^2}{N_1^2}\left(\frac{\lambda_4}{\lambda_1}\right)^k + \frac{(\alpha_1\alpha_2 - \gamma_1\gamma_2)^2}{N_1^2 N_2^2}\left(\frac{\lambda_2}{\lambda_1}\right)^k +$$

$$+ \frac{(\alpha_1\alpha_3 - \gamma_1\gamma_3)^2}{N_1^2 N_3^2}\left(\frac{\lambda_3}{\lambda_1}\right)^k + \frac{(\alpha_1^2 - \gamma_1^2)^2}{N_1^4}.$$

*The correlation function for the same chain and different chains spins on the open strip in the thermodynamic limit, respectively:*

$$G_{i,j} = \frac{(\alpha_1\alpha_2 - \gamma_1\gamma_2)^2}{N_1^2 N_2^2}\left(\frac{\lambda_2}{\lambda_1}\right)^k + \frac{(\alpha_1\alpha_3 - \gamma_1\gamma_3)^2}{N_1^2 N_3^2}\left(\frac{\lambda_3}{\lambda_1}\right)^k + 2\frac{\beta_1^2}{N_1^2}\left(\frac{\lambda_4}{\lambda_1}\right)^k,$$

$$G_{i,j} = \frac{(\alpha_1\alpha_2 - \gamma_1\gamma_2)^2}{N_1^2 N_2^2}\left(\frac{\lambda_2}{\lambda_1}\right)^k + \frac{(\alpha_1\alpha_3 - \gamma_1\gamma_3)^2}{N_1^2 N_3^2}\left(\frac{\lambda_3}{\lambda_1}\right)^k - 2\frac{\beta_1^2}{N_1^2}\left(\frac{\lambda_4}{\lambda_1}\right)^k.$$

## III. MAIN SPECIAL CASE: HAMILTONIAN WITH INTERACTIONS OF EVEN NUMBER OF SPINS

### A. Case description

We introduce a model with interactions of an even number of spins. Hamiltonian will be represented as:

$$\mathcal{H}^m = -J_1\left(\sigma_0^m\sigma_0^{m+1} + \sigma_1^m\sigma_1^{m+1}\right) - \frac{J_2}{2}\left(\sigma_0^m\sigma_1^m + \sigma_0^{m+1}\sigma_1^{m+1}\right) -$$

$$- J_3\left(\sigma_0^m\sigma_1^{m+1} + \sigma_0^{m+1}\sigma_1^m\right) - J_4\sigma_0^m\sigma_1^m\sigma_0^{m+1}\sigma_1^{m+1}. \quad (20)$$

Due to the specific form of the Hamiltonian (20), the transfer matrix will have central-symmetric structure:

$$\Theta_2 = \begin{pmatrix} a & b & b & c \\ b & d & f & b \\ b & f & d & b \\ c & b & b & a \end{pmatrix}. \quad (21)$$

Eigenvalues of matrix (21) are:

$$\lambda_{1,2} = \frac{1}{2}\left(a + c + d + f \pm \sqrt{(a + c - d - f)^2 + 16b^2}\right), \quad (22)$$

$$\lambda_3 = a - c,$$

$$\lambda_4 = d - f.$$

The main theorem (theorem 2) remains correct in the considered special case, where $\lambda_1 = \lambda_{max}$ is calculated using the first formula (22). The diagonalizing matrix has the form:

$$Q_2 = \begin{pmatrix} \frac{1}{M^+} & \frac{1}{M^-} & \frac{1}{\sqrt{2}} & 0 \\ \frac{\varphi^+}{M^+} & \frac{\varphi^-}{M^-} & 0 & \frac{1}{\sqrt{2}} \\ \frac{\varphi^+}{M^+} & \frac{\varphi^-}{M^-} & 0 & -\frac{1}{\sqrt{2}} \\ \frac{1}{M^+} & \frac{1}{M^-} & -\frac{1}{\sqrt{2}} & 0 \end{pmatrix},$$

where expressions are:

$$\varphi^+ = \frac{-a - c + d + f + \sqrt{(a + c - d - f)^2 + 16b^2}}{2b},$$

$$\varphi^- = \frac{-a - c + d + f - \sqrt{(a + c - d - f)^2 + 16b^2}}{2b},$$

$$M^\pm = \sqrt{2 + 2(\varphi^\pm)^2}.$$

We provide the proof of this statement below (section 4).

### B. Correlation functions and correlators

In the considered special case we can apply the theorem 3 and derive corollary theorem.

**Theorem 4.** *For a two-chain open model with interactions of an even number of spins where $L \to \infty$:*

*The average spin value of $\sigma_i$:*

$$\langle\sigma_i\rangle = 0.$$

*The correlation function for the same chain and different chains spins on the open strip in the thermodynamic limit, respectively:*

$$G_{i,j} = \frac{(2-\sqrt{2})^2}{4(M^+)^2}\left(\frac{\lambda_3}{\lambda_1}\right)^k + 2\frac{(\varphi^+)^2}{(M^+)^2}\left(\frac{\lambda_4}{\lambda_1}\right)^k,$$

$$G_{i,j} = \frac{(2-\sqrt{2})^2}{4(M^+)^2}\left(\frac{\lambda_3}{\lambda_1}\right)^k - 2\frac{(\varphi^+)^2}{(M^+)^2}\left(\frac{\lambda_4}{\lambda_1}\right)^k,$$

*where $k=|j-i|$.*

*Remark 1.* In particular, with a specific relationship between the parameters [6] in the Hamiltonian (20), we obtain the gonihedric model.

### C. Ground states and correlation length in main special case

Transfer matrix (21) has five unique elements, representing five unique states of one plaquette of the lattice. We can write them as

$$a = e^{\beta(2J_1 + J_2 + 2J_3 + J_4)}, \quad (23)$$

$$b = e^{\beta(-J_4)}, \quad (24)$$

$$c = e^{\beta(-2J_1 + J_2 - 2J_3 + J_4)}, \quad (25)$$

$$d = e^{\beta(2J_1 - J_2 - 2J_3 + J_4)}, \quad (26)$$

$$f = e^{\beta(-2J_1 - J_2 + 2J_3 + J_4)}, \quad (27)$$

$$\beta = T^{-1}.$$

For $J_1>0$ we can obtain a ground states diagram in new parameters $J_2'=J_2/J_1$, $J_3'=J_3/J_1$ and $J_4'=J_4/J_1$ by minimizing the Hamiltonian corresponding to each state. Similar expressions for $J_1<0$ can be found [4].

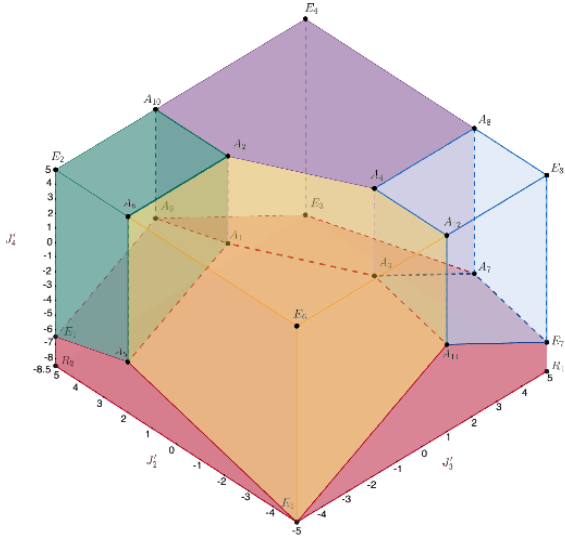

*Figure 2. Ground states diagram.*

Coordinates of the vertices are represented in the table 2. Each state (23)-(27) is shown using colors as shown (Figure 3). The domains can be continued throughout the whole $\mathbb{R}^3$.

The boundary surfaces are of special interest, because the correlation length behaves in an unusual way. One can observe that at these points, frustration points, the inverse correlation length approaches a non-zero constant as temperature approaches zero. Also, at these points such previously mentioned physical quantities as free energy, internal energy, specific heat capacity and entropy also change their behaviors. These phenomena are shown as examples in figures 3 – 6. The inverse correlation length can be written as [5]:

$$\xi^{-1} = \ln\left(\min_{\lambda_i \neq \lambda_{max}} \left|\frac{\lambda_{max}}{\lambda_i}\right|\right).$$

*Table 2. Coordinates of the boundary points of the regions of the ground states shown in figure 2.*

| Point | Coordinates | Point | Coordinates |
|---|---|---|---|
| $A_1$ | $(-1,2,-1)$ | $A_2$ | $(-1,2,5)$ |
| $A_3$ | $(1,-2,-1)$ | $A_4$ | $(1,-2,5)$ |
| $A_5$ | $(-5,2,-5)$ | $A_6$ | $(-5,2,5)$ |
| $A_7$ | $(5,-2,-5)$ | $A_8$ | $(5,-2,5)$ |
| $A_9$ | $(-1,5,-5/2)$ | $A_{10}$ | $(-1,5,5)$ |
| $A_{11}$ | $(1,-5,-5/2)$ | $A_{12}$ | $(1,-5,5)$ |
| $E_1$ | $(-5,5,-13/2)$ | $E_2$ | $(-5,5,5)$ |
| $E_3$ | $(5,5,-17/2)$ | $E_4$ | $(5,5,5)$ |
| $E_5$ | $(-5,-5,-17/2)$ | $E_6$ | $(-5,-5,5)$ |
| $E_7$ | $(5,-5,-13/2)$ | $E_8$ | $(5,-5,5)$ |
| $R_1$ | $(5,-5,-17/2)$ | $R_2$ | $(-5,5,-17/2)$ |

*Table 3. Colors of ground states on diagram (Figure 2).*

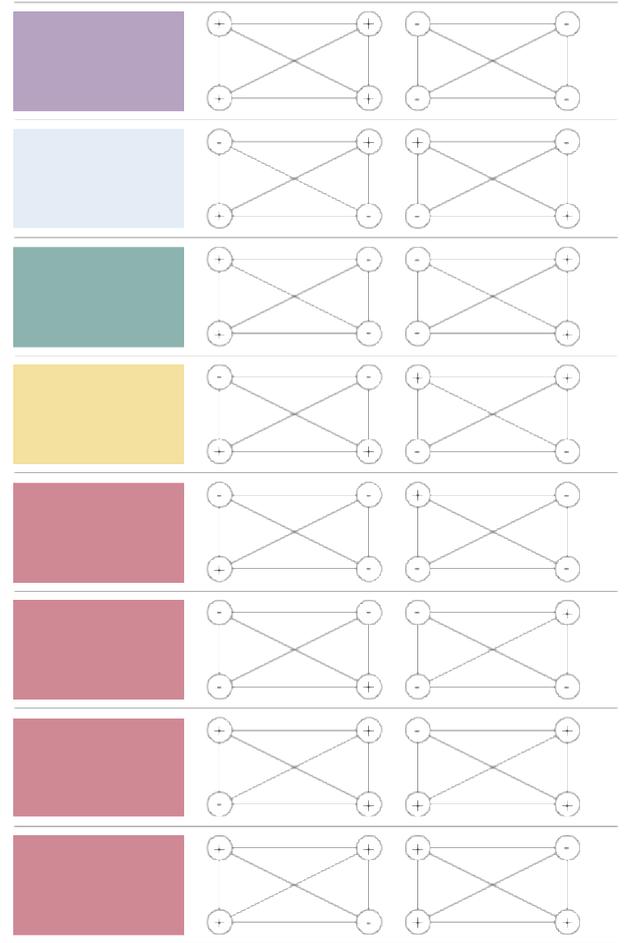

As the first example, where three ground states coincide, the point $A_1$ is chosen. Plots (Figures 3,4) show the behavior at this frustration points.

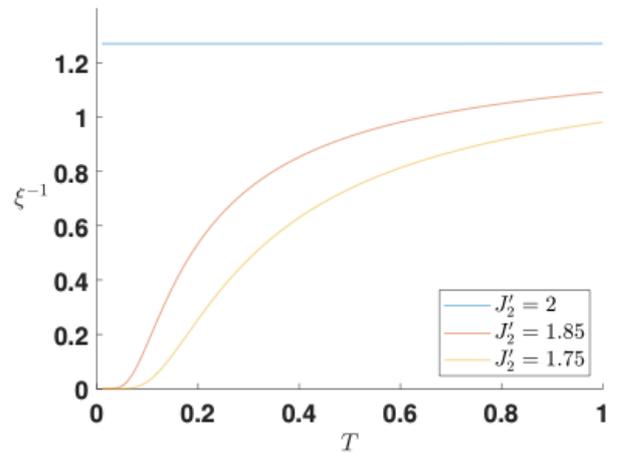

*Figure 3. Inverse correlation length plot in the low-temperature region at the point $A_1$, where $J_3'=-1$, $J_4'=-1$, $T \in [0.01;1]$, $J_2'=2, 1.85, 1.75$.*

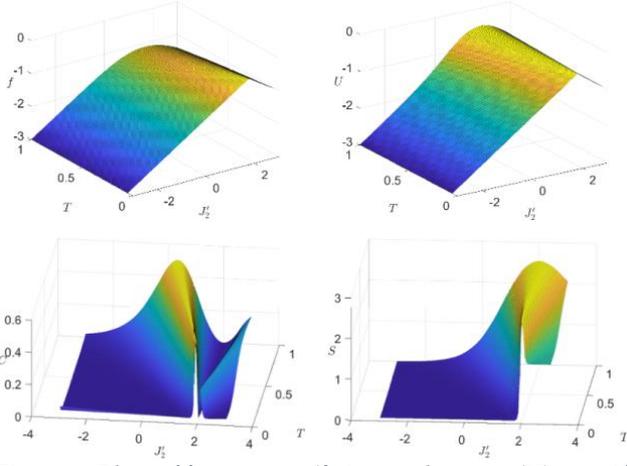

*Figure 4. Plots of free energy (f), internal energy (U), specific heat (c) and entropy (S) at the point $A_1$, $J_3'=-1$, $J_4'=-1$, $T \in [0.01;1]$, $J_2' \in [-3;3]$.*

As the second example, where two ground states coincide, the point $E_3$ is chosen. Plots (Figures 5,6) show behavior of quantities at this frustration point.

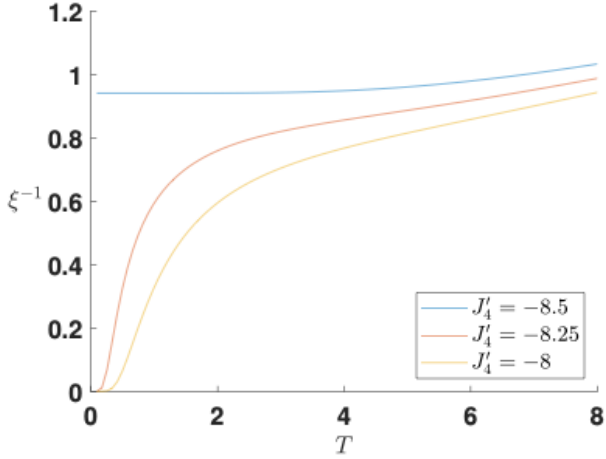

*Figure 5. Inverse correlation length plot in the low-temperature region at the point $E_3$, where $J_2'=5$, $J_3'=5$, $T \in [0.01;8]$, $J_4'= -8.5, -8.25, -8$.*

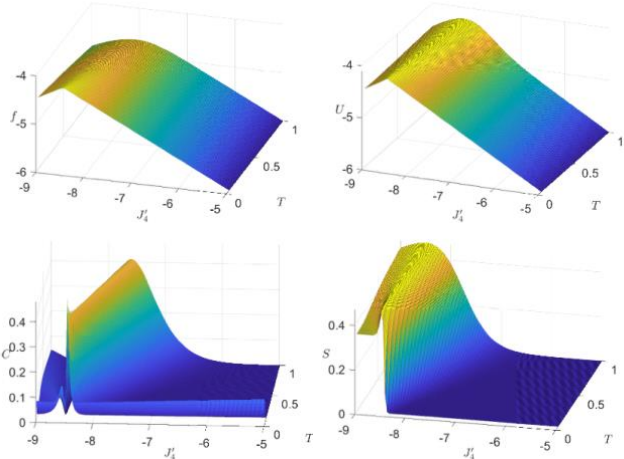

*Figure 6. Plots of free energy (f), internal energy (U), specific heat (c) and entropy (S) at the point $E_3$, $J_2'=5$, $J_3'=5$, $T \in [0.01;1]$, $J_4' \in [-9;-5]$.*

## IV. THEOREM PROOFS

### A. Finding eigenvalues of transfer matrix in general case

We can consider the commuting permutation matrix for matrix (4):

$$P = \begin{pmatrix} 1 & 0 & 0 & 0 \\ 0 & 0 & 1 & 0 \\ 0 & 1 & 0 & 0 \\ 0 & 0 & 0 & 1 \end{pmatrix}.$$

It has one eigenvector– $(0,1,-1,0)^T$ - corresponding to $\lambda_4$, three other eigenvectors have equal second and third components, so we obtain the matrix:

$$\begin{pmatrix} a & 2b & c \\ b & d+f & g \\ c & 2g & h \end{pmatrix},$$

and its characteristic polynomial gives us the exact three remaining eigenvalues. To find them, it is necessary to solve the cubic equation, roots are found using formula in theorem 1, where $\lambda_1$ is positive and has the greatest absolute value, according to the Perron-Frobenius theorem.

### B. Finding eigenvalues of transfer matrix in case of Hamiltonian with even number of interactions

Transfer matrix in this case is commuting with the permutation matrix:

$$P_2 = \begin{pmatrix} 0 & 0 & 0 & 1 \\ 0 & 0 & 1 & 0 \\ 0 & 1 & 0 & 0 \\ 1 & 0 & 0 & 0 \end{pmatrix},$$

which has eigenvectors:

$$\bar{v}_1 = \begin{pmatrix} 1 \\ 0 \\ 0 \\ 1 \end{pmatrix}, \bar{v}_2 = \begin{pmatrix} 0 \\ 1 \\ 1 \\ 0 \end{pmatrix}, \bar{v}_3 = \begin{pmatrix} 1 \\ 0 \\ 0 \\ -1 \end{pmatrix}, \bar{v}_4 = \begin{pmatrix} 0 \\ 1 \\ -1 \\ 0 \end{pmatrix}.$$

So the eigenvectors of the matrix (21) will be like:

$$\bar{x}_{1,2} = \begin{pmatrix} x_1 \\ x_2 \\ x_2 \\ x_1 \end{pmatrix}, \bar{x}_{3,4} = \begin{pmatrix} x_1 \\ x_2 \\ -x_2 \\ -x_1 \end{pmatrix}.$$

So we introduce two matrices to help us find the eigenvalues:

$$\tau' = \begin{pmatrix} a+c & 2b \\ 2b & d+f \end{pmatrix},$$

$$\tau'' = \begin{pmatrix} a-c & 0 \\ 0 & d-f \end{pmatrix}.$$

Eigenvalues of transfer matrix are eigenvalues of $\tau'$ and $\tau''$, respectively:

$$\lambda_{1,2} = \frac{1}{2}\left(a+c+d+f \pm \sqrt{(a+c-d-f)^2 + 16b^2}\right),$$

$$\lambda_3 = a-c,$$

$$\lambda_4 = d-f.$$

Eigenvalue $\lambda_1$ is positive and has the greatest absolute value, according to the Perron-Frobenius theorem.

V. Conclusion

In this paper, the exact expressions for the partition function of a two-chain Ising model of finite length with an external field, double, triple and quadruple interactions, free energy, internal energy, specific heat, susceptibility, magnetization and entropy in the thermodynamic limit $L \to \infty$ are obtained using the transfer matrix method.

In section 2 three theorems are formulated. Theorem 2 provides analytic formulas to calculate the partition function, eigenvalues of the transfer matrix (4) and, especially, positive and greatest in absolute value eigenvalue $\lambda_{max}$. The roots of the quartic characteristic polynomial are found as well as the system of eigenvectors of the transfer matrix, using an additional matrix. Eigenvalue $\lambda_{max}$ is used in the main theorem (theorem 1) and that provides exact formulas of physical quantities in the thermodynamic limit which are of particular interest. Then formulas for the correlation functions in the thermodynamic limit are formulated for spins located on the same chain and different chains.

The solution of a model with interactions of an even number of spins is also found. Three theorems from second chapter can be applied in this special case and the same partition function formulas can be derived. By minimizing the Hamiltonian, corresponding to each state, the structure of ground states is obtained, a phase diagram is shown and described for the case of positive parameter $J_1$. Then graphs of free energy, internal energy, specific heat capacity and entropy are shown with the graph of inverse correlational length near two frustration points $A_1$ and $E_3$ as examples.

Acknowledgment

Authors are grateful to G. Skvortsov for helpful comments and discussion.